\documentclass[prb,twocolumn,floatfix,showpacs,superscriptaddress]{revtex4-1}
\usepackage{graphicx,epsfig,color}
\usepackage[%  
    colorlinks=true,
    pdfborder={0 0 0},
    allcolors=blue,
    breaklinks=true
]{hyperref}

\usepackage{threeparttable}
\usepackage{breakcites}
\usepackage[utf8]{inputenc}
\begin{document}

\title{Use of local density approximation within range separated hybrid exchange-correlation functional to investigate Pb doped SnO$_2$ as an electron transport layer}

\author{Veysel \c{C}elik}\email{veysel3@gmail.com}
\affiliation{Department of Mathematics and Science Education, Siirt University, Siirt 56100, Turkey}

\date{\today}

\begin{abstract}
In this study, the structural, electronic and optical properties of Pb doped rutile SnO$_2$ were investigated using the range separated hybrid exchange-correlation functional method. In the calculations, LDA functional was used instead of PBE functional. The electronic structure of SnO$_2$ obtained by this method is quite compatible with the experimental data. The SnO$_2$ has an important usage area in optoelectronic devices due to its transparent and conductive nature. One of these important areas is the use of SnO$_2$ as an electron transport layer (ETL) in perovskite solar cells. Therefore, the energy level of the conduction band of the SnO$_2$ is important. In the Pb doped SnO$_2$ cases, the band gap narrows as the Pb doping rate increases. The bandgap of SnO$_2$ can be narrowed from 3.60 eV to 3.02 eV with a \%12.5 Pb doping ratio, and this narrowing is proportional to the amount of Pb. The calculation results obtained in this study show that the decrease in the energy level of the bottom of the conduction band plays an important role in the narrowing of the band gap and there is no significant change in the energy level of the top of the valence band. Due to this effect of the Pb atom, the energy level of the conduction band can be adjusted by using the doping ratio of the Pb atom and the band gap can be narrowed in a controlled manner. With the Pb doping, the energy levels of the SnO$_2$ ETL can be adjusted in a range according to the type of perovskite used in solar cell. In addition, the doping with Pb does not create electron traps in the band gap, which is important in the transport process of electrons.
\end{abstract}


\maketitle
                                    
\section{Introduction}
Due to its combination of high electrical conductivity and optical transparency, transparent conductive oxides (TCO) are important for solar cell. \cite{Klein2010,Zhang2018} Among the TCO materials, tin dioxide (SnO$_2$) has excellent electrical, optical and electro-chemical properties. Tin dioxides have been widely used for solar panels\cite{Jiang2016,Ellmer2012} and touch screens.\cite{KIKUCHI2002} For these mentioned applications, it is important to be able to adjust the bandgap that SnO$_2$ has. The SnO$_2$ has a direct band gap whose width is about 3.6 eV(Ref.\citenum{Godinho2009}) which is in the ultraviolet region. Hence, SnO$_2$ is transparent under sunlight, and this is a useful feature for solar cells. Perovskite-based solar cells (PSC) are one of the important areas where SnO$_2$ is used as an electron transport layer (ETL).\citep{Zhang2018,CHEN2019144} Among the metal oxides, TiO$_2$ is the most commonly used ETL material in PSC devices. Due to its proper band gap and high transmittance, TiO$_2$ is preferred for PSCs, but its electron mobility is shorter than perovskite.\citep{CHEN2019144} Compared to TiO$_2$, SnO$_2$ has high electron mobility, deep CB level, and usability for flexible solar cells \citep{CHEN2019144} However, there are many types of perovskites, the energy levels of ETL materials must match the perovskite materials, and this can be done by fine-tuning the energy level of the CB. One of the most important ways to change the energy levels of the CB and valence band (VB) is doping with the foreign atoms. 

Many experimental studies have been conducted on how the structural, electronic and optical properties of SnO$_2$ have changed when it is doped with foreign atoms.\cite{Chetri2013,Chen2011,WANG2009,walle2011,XIAO2009,FAKHIMLAMRANI2011,AHMED20111,AZAM201283,Adhikari2008,Bouaine2007,Hays2005} The Ni doping rate of \%9 narrows the band gap from 3.90 eV to 3.31 eV and shortens the lattice parameters according to the pure case.\cite{AHMED20111} In case of doping with Mn, the band gap expands as the doping rate increases, and with \%15 the band gap increases from 3.71 eV to 4 eV.\citep{AZAM201283} In case of Cu doping, the bandwidth expands from 3.93 eV to 4.00 eV, with a \%5 doping rate.\citep{Chetri2013} On the other hand, the Cu-doped SnO$_2$ quantum dots samples exhibited enhanced absorption capability in the visible light region and the band gap decreased to about 2.2 eV by the increasing concentration of Cu.\cite{BABU2017330} In the experimental study with Pb doping, it was found that the \%15 doping rate narrowed the bandwidth from 3.64 eV to 2.87 eV.\cite{SARANGI201816} The narrowing is proportional to the amount of Pb. This property of the Pb atom can be used for SnO$_2$ ETLs. However, the mechanism of this narrowing should be explained in more detail. Theoretical investigations are important in this sense. In the density functional theory (DFT) frame, many studies have been carried out on the changes in the structural and electronic properties of impurity-doped SnO$_2$.\cite{Chen2011,WANG2009,walle2011,XIAO2009,FAKHIMLAMRANI2011} The general problem in theoretical studies is to calculate the band gap closer to the experimental data. This is one of the focal points in this study. The main purpose of this study is to reveal how the Pb doping narrows the band gap and what kind of structure it forms in the material when doped with more than one Pb atom. These obtained data will shed light on the fine-tuning of the energy levels of SnO$_2$ with Pb doping for ETLs.

In this study, structural, electronic and optical properties of Pb doped SnO$_2$  were investigated. Pb is in the same group as Sn and has a larger ionic radius. In this present work,  screened Coulomb potential hybrid DFT calculations was used to investigate the modifications of the electronic properties of SnO$_2$ induced by one Pb doping (Pb$_1$), two Pb doping (Pb$_2$) and three Pb doping (Pb$_3$) cases and their effects on the corresponding imaginary part of the dielectric constant. In the calculations, LDA functional was used instead of PBE in HSE06 hybrid functional. Before investigating the Pb doping cases, necessary tests were carried out and it was predicted that the use of mixing the LDA functional correctly predicted the electronic structure and would give correct results for the Pb atom in the same group with the Sn atom. Formation energies have been calculated as a function of oxygen chemical potential to compare thermodynamical stability of the doped structures.

\begin{figure}[h]
    \centering
    \includegraphics[width=0.3\textwidth]{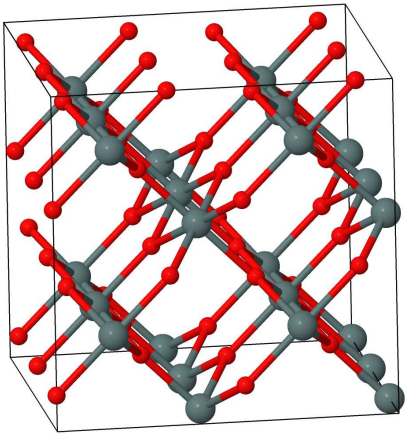}
    \caption{Sn$_{24}$O$_{48}$ supercell. Red small and large gray
balls presented O and Sn, respectively.}
    \label{fig1}
\end{figure}

\section{Computational Method}
The spin-polarized hybrid density functional theory (DFT) calculations have been performed based on the projector-augmented wave (PAW)\cite{Blochl,Kresse2} method as implemented in the Vienna ab-initio simulation package (VASP).\cite{Kresse1,Kresse3} The exchange-correlation effects have been taken into account by employing the range separated hybrid HSE functional.\cite{HSE03,Paier2006} In the calculations, the pseudopotentials in the Local Density Approach (LDA)\cite{Kohn1965,Kresse3,LDACA} framework were used. The Sn potential which treats the 4d semicore states as valence states were used for bulk rutile SnO$_2$ calculation. On the other hand, due to the high computational cost, the potential that treats the 4d semicore states as core states was used in the doping case calculations. According to the tests performed in this study, using these two types of potentials in HSE calculation gives similar results for the structure of VB, CB and crystal.

The lack of a proper self-interaction correction (SIC) leads to 
the well-known band gap underestimation by the standard DFT 
exchange-correlation functionals, such as Perdew–Burke–Ernzerhof exchange-correlation
functional (PBE).\cite{PBE1996}
On the other hand, Hartree-Fock (HF) formalism has well-defined Coulomb
direct and exchange terms canceling each other for the zero momentum 
components avoiding self-interaction of charges. In order to benefit 
from this, modern hybrid DFT functionals partially admix the nonlocal exact 
exchange energy with the semilocal PBE exchange energy. 

The hybrid HSE functional treats the exchange energy as composed of long-range 
(LR) and short-range (SR) parts with a range separation parameter $\omega$ and 
mixes the exact exchange with the PBE exchange at the short-range by a mixing 
factor of $a=0.25$ such that,\cite{HSE03,Paier2006}
\[
E_{\textbf{\tiny X}}^{\textrm{\scriptsize HSE}}=
a E_{\textbf{\tiny X}}^{\textrm{\scriptsize HF,SR}}(\omega)+
(1-a)E_{\textbf{\tiny X}}^{\textrm{\scriptsize PBE,SR}}(\omega)+
E_{\textbf{\tiny X}}^{\textrm{\scriptsize PBE,LR}}(\omega)\,.
\]
The correlation term of the XC energy is taken from standard PBE correlation energy.\cite{PBE1996} However, in this present work, with the same formula, hybrid functional mix local exchange potentials in the LDA with the exact nonlocal Hartree–Fock exchange potential. At the same time, the correlation energy is described within the LDA. The exact exchange contribution was determined to take into account the experimental data. In this context,  the exact exchange contribution of \%29 described closest to experimental band gap value of SnO$_2$ for the rutile phase. This rate is close to the standard rate of \%25.

In this work, 72-atom rutile bulk supercell containing 24 Sn atoms and 48 O atoms used for doped systems, as shown in Fig. 1. Supercells arised from 2$\times$2$\times$3 replication of the rutile unit cell of SnO$_2$. The Pb atom was substituted for Sn atoms. For geometry optimizations and electronic-structure, the Brillouin zones were sampled with 2$\times$2$\times$2 Monkhorst-Pack~\cite{mp} $k$-point grids for 72 atoms supercells. Plane wave basis set was used to expand the wavefunctions up to a kinetic energy cutoff value of 450 eV. The fine FFT grids with high precision settings were used throughout the calculations. Atomic positions and cell parameters were optimized until residual forces were below 0.015 eV/\AA.

For the qualitative description of interatomic charge distributions, Bader analysis based on atom in molecule (AIM) theory used. Local charge depletion/accumulation can be computed by integrating Bader volumes around atomic sites. These volumes are partitions of the real space cell delimited by local zero-flux surfaces of charge density gradient vector field. Charge states of atomic species (see Table II) were calculated  using a grid based decomposition algorithm developed by Henkelman’s group.\cite{Tang_2009}

\begin{table}[h]%[htb]
\label{table1}
\begin{threeparttable}
\caption{The comparison of computational and experimental data for rutile bulk SnO$_2$.
Latice parameters and band gaps are in angstroms and eV, respectively.}
\begin{ruledtabular}
\begin{tabular}{lcccccc}
\multicolumn{1}{l}{Functional}& \multicolumn{2}{l}{Latice parameters (\AA)}&\multicolumn{1}{c}{Band gaps (eV)}\\[1mm]\hline
& a & c &\\[1mm] \hline
LDA\tnote{a}& 4.73 & 3.20& 1.08\\
GGA\tnote{b}  & 4.83 & 3.24 & 0.65 \\
GGA+U(4 eV)\tnote{c}  & 4.73 & 3.16 & 1.93 \\
HSE06\tnote{d}& 4.76 & 3.19 & 2.96 \\
HSE03+G$_0$W$_0$\tnote{a}  & - & -& 3.65 \\
This Work   & 4.70 & 3.16 & 3.60 \\
Experimental  & 4.73\tnote{e} & 3.18\tnote{e} & 3.60\tnote{f} \\
\end{tabular}
\end{ruledtabular}
\begin{tablenotes}
        \item[a] Reference\citenum{varley2011}.
        \item[b] Reference\citenum{Borges2010}.
        \item[c] Reference\citenum{Stashans2014}.
        \item[d] Reference\citenum{varley2009}.
        \item[e] Reference\citenum{Bolzan1997}.
        \item[f] Reference\citenum{Godinho2009}.
    \end{tablenotes}
\end{threeparttable}
\end{table}

\begin{table}[h]%[htb]
\caption{Average charge states (\textit{e}) of dopants and their adjecent Sn and O atoms from Bader analysis.}
\begin{ruledtabular}
\begin{tabular}{cccccccc}
\multicolumn{1}{c}{Structure}&\multicolumn{1}{c}{Pure}&\multicolumn{1}{c}{Pb$_1$ }&\multicolumn{1}{c}{Pb$_2$ }&\multicolumn{1}{c}{Pb$_3$ }\\[1mm]\hline
Sn$_{n1}$ &+2.42&+2.41&+2.42&+2.42\\%\cite{haffad}&1.48\cite{haffad} \\
Sn$_{n2}$ &+2.42&+2.40&+2.41&+2.42\\%\cite{haffad}&1.48\cite{haffad} \\
O$_{n1}$ &-1.20&-1.14&-1.14&-1.15\\
O$_{n2}$ &-1.20&-1.14&-1.14&-1.13\\
O$_{nn}$  &-&-&-1.06&-1.06\\
Pb$_{11}$  &-&+2.01&-&-\\
Pb$_{21}$   &  -& -&+2.02&-\\
Pb$_{22}$   &-& -&+1.96&-\\
Pb$_{31}$   &-& -&-&+1.98\\
Pb$_{32}$   &-& -&-&+2.02\\
Pb$_{33}$   &-& -&-&+2.01\\
\end{tabular}
\end{ruledtabular}
\label{table2}
\end{table}

\section{Results \& Discussion}
Previous theoretical studies show that HSE functional is 
useful to get electronic band gap related features of periodic and finite 
physical systems reasonably accurate to be comparable with 
experiments.\cite{Henderson2011,Celik2012,Vines2} However, for rutile SnO$_2$, in the calculations made by HSE06 method, which were used the default \%25 mixing rate, the bandgap was calculated 2.96 eV (Ref.\citenum{varley2009}). This value is less than the experimental value of 3.6 eV(Ref.\citenum{Godinho2009}). \citet{Behtash2015}, Increased the mix ratio to 33 percent to approximate the experimental band gap value in the HSE06 calculation. In another study,  using first-order perturbation theory based on an initial electronic structure from HSE03, the bandgap was calculated 3.65 eV.\cite{varley2011} As an alternative method, in this present work, with a \%29 mixing rate, the calculated value for the band gap is 3.60 eV. The obtained DOS pattern for bulk rutile SnO$_2$ is shown in Fig 2. The averaged d states is located 21.1 eV below the valence band maximum (VBM), which is good agreement with the reported experimental values of 21.1 eV (Ref.\citenum{Sherwood1990}) and 21.5 eV (Ref.\citenum{Themlin1992}). The PBE and LDA calculations underestimate the binding energy of the averaged d states and their positions are 19.9 eV and 19.7 eV below the VBM for PBE and LDA, respectively. As can be seen in Fig. 2, the calculated valance band (VB) width is 8.7 eV and this value is close to the experimental value of 8.5 eV from resonant photoelectron spectroscopy.\cite{Haeberle2016} In the VB, O p energy levels are dominant. There are Sn d energy levels below the top of the VB. In the lower parts of the conduction band (CB), Sn s-O p energy levels are dominant. The characteristics of the electronic structure obtained with the calculations made in this study are compatible with the standard DFT and hybrid DFT studies that have been done before.\citep{Behtash2015,varley2011,Yuhua2008,Mishra1995,ZAINULLINA2007280,Hamad2009} By comparison, LDA can describe rutile SnO$_2$ better than GGA-PBE. One of the aims of this present work is to use this feature of the LDA.

\begin{figure}[h]
    \centering
    \includegraphics[width=0.5\textwidth,trim={0 0 0 0},clip]{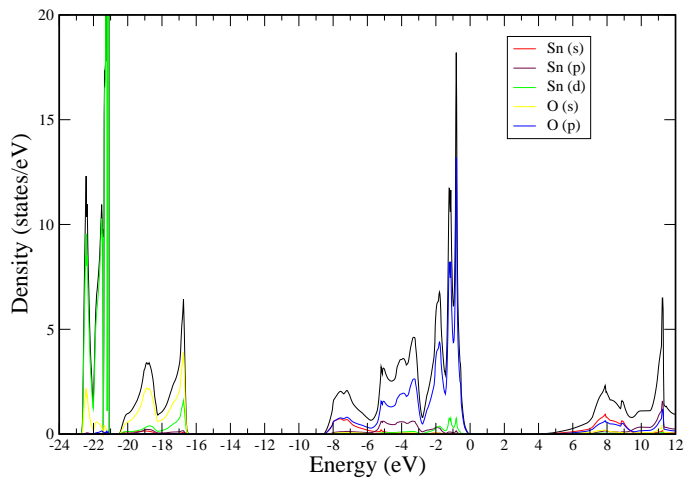}
    \caption{Densities of states (DOS) of bulk rutile SnO$_2$.}
    \label{fig2}
\end{figure}

In terms of structural features, lattice parameters can be calculated very close to experimental value with LDA functional, but relatively higher with GGA-PBE functional (See Table I). With the default mixing parameter of \%25 in HSE functional, by correcting the overestimate in the calculation made with GGA-PBE functional, the lattice parameters can be calculated close to the experimental value, but, on the other side, the value of the band gap is calculated lower than it should be. In this present work, calculated lattice parameters are a = 4.70 \AA\ and c = 3.16 \AA, which are slightly underestimate compared with the experimental values\cite{Bolzan1997} (a = 4.73 \AA\ and c = 3.18 \AA). As mentioned earlier, using standard LDA, the lattice parameter can be calculated very close to experimental data. Increasing the amount of the exact exchange contribution causes the lattice parameter to shorten while expanding the bandgap. However, these calculated values are acceptable.

In this work, the dopant formation energies have been calculated using,
\[
E_{f}=E_{\rm doped}-E_{\rm pure}-n\mu_{\rm Pb}+n\mu_{\rm Sn},
\]
where E$_{\rm doped}$ and E$_{\rm pure}$ are the total energies of doped 
and pure supercells, while $\mu_{\rm Pb}$ and $\mu_{\rm Sn}$ are the chemical potentials of the Pb and Sn species, respectively. The integer $n$ gives the number of Pb cations. In thermodynamical equilibrium with the rutile phase, $\mu_{\rm Sn}$ and $\mu_{\rm O}$ must 
satisfy the relation $\mu_{\rm SnO_{2}}=\mu_{\rm Sn}+2\mu_{\rm O}$. The amount of Sn and O in a growth environment influences their chemical potentials. High(low) values of $\mu_{\rm O}$ correspond to O-rich(-poor) conditions and 
can also be interpreted as Sn-poor(-rich) conditions from the equilibrium 
relation. Under O-rich conditions, $\mu_{\rm O}$ is the half of the energy 
of an O$_2$ molecule $(E_{\rm O_2})$, and $\mu_{Sn}$ is obtained through 
the condition $\mu_{\rm Sn}=\mu_{\rm SnO_{2}}-E_{\rm O_2}$.
 Under Sn-rich conditions, $\mu_{\rm Sn}$ is the energy of a Sn atom in its bulk unit cell 
($\mu_{\rm Sn}^{\rm bulk}$) and $\mu_{\rm O}$ is calculated from the 
equilibrium restriction by $\mu_{\rm O}=\frac{1}{2}(\mu_{\rm SnO_{2}}-\mu_{\rm Sn})$. 
The chemical potentials of the dopants are extracted from their naturally 
occurring phases. $\mu_{\rm Pb}$ is the energy of a Pb atom in its bulk unit cell 
($\mu_{\rm Pb}^{\rm bulk}$). Calculated dopant formation energies are given in Fig. 8. In this study, the formation enthalpy calculated for bulk rutile SnO$_2$ is -6.1 eV and this value is very close to the experimental value of about -6.0 eV(Ref.\citenum{CRC2009}).

\begin{table}[h]%[htb]
\caption{Comparison of the computational result of the band gaps according to the doping cases. The band gap energies are in eV.}
\begin{ruledtabular}
\begin{tabular}{cccccccc}
\multicolumn{1}{c}{Structure}&\multicolumn{1}{c}{Band Gap (eV)}&\multicolumn{1}{c}{Transition}\\[1mm]\hline
Pure  & 3.60& $\Gamma \rightarrow \Gamma $\\%\cite{haffad}&1.48\cite{haffad} \\
Pb$_1$ &3.39 & $\Gamma \rightarrow \Gamma $ \\
Pb$_2$ &3.17& $\Gamma \rightarrow \Gamma $\\
Pb$_3$ &3.02& $\Gamma \rightarrow \Gamma $\\
\end{tabular}
\end{ruledtabular}
\label{table3}
\end{table}
\begin{figure*}[t]
    \centering
    \includegraphics[width=0.9\textwidth]{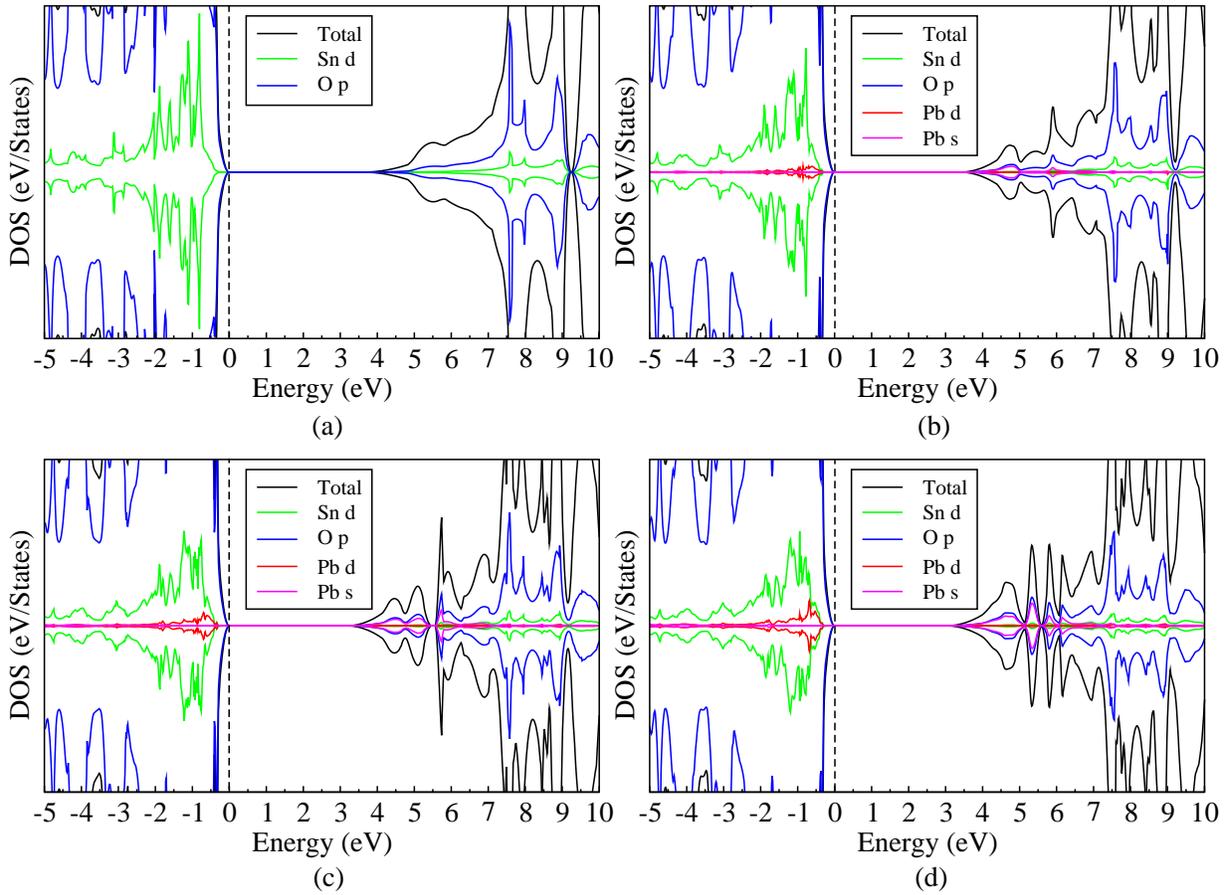}
    \caption{Densities of states (DOS) of (a) pure, (b) Pb$_1$, (c) Pb$_2$ and (d) Pb$_3$ structures. Dotted line indicates the Fermi energy.}
    \label{dos}
\end{figure*}

\textbf{\textit{One Pb doping(Pb$_1$)}}: According to the calculations in this study, in accordance with experimental data,\cite{SARANGI201816} it is more convenient to replace the Pb atom with an Sn atom in terms of energy. In this structure, there are 6 adjacent O atoms to a Pb atom. As will be seen in Fig. 4, the bond length of these 6 oxygen atoms with the Pb atom is 2.13\AA\ for the 4 cross-located oxygen atoms and 2.10\AA\ for the other 2 oxygen atoms. For Sn atoms shown in Fig. 4, the bond length between Sn and O atoms was calculated as about 2.02\AA\ in the Pb doped SnO$_2$. In pure SnO$_2$, the Sn-O bond length is 2.03\AA. There is no noticeable change in Sn-O bond length compared to the pure case. Since Pb atom has a larger ionic radius than Sn atom, the Pb atom pushes the adjacent O atoms outward in a small amount. At the same time, the angle between the O atoms in the O-Sn-O chain opens according to the pure case so that the Sn-O bond lengths do not change much. Pb doping does not distort the crystal structure considerably, and the distortion that occurs is local. This result is consistent with the X-ray diffraction results obtained in an experimental study.\cite{SARANGI201816}

\begin{figure}[h]
    \centering
    \includegraphics[width=0.3\textwidth]{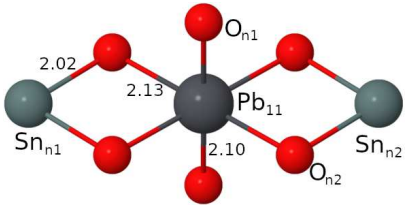}
    \caption{Relaxed structure of the one Pb dopant inside the rutile SnO$_2$. The Bond lengths are in angstroms.}
    \label{s1pb}
\end{figure}
The density of states (DOS) plots in Fig. 3(b) show the electronic structure of the single Pb doping. Impurity incorporation hasn't induced a local empty state in the energy gap between the VB and the CB. The clearance of the bandgap indicates that there are no electron traps in the bandgap. The electron traps can increase the possibility of electron-hole recombination before the electron is transferred to the conduction band. As can be seen in Table III, the band gap is direct ($\Gamma \rightarrow \Gamma$) and has a width of 3.39 eV. According to the pure case, single Pb doping narrows the band gap. The imaginary parts of dielectric constant values versus photon energy of Pb doped SnO$_2$ with different doping levels are shown in Fig. 5. The imaginary part ($\varepsilon_2$) of the dielectric constant depends mainly on the coefficient of extinction, which depends on the variation of the absorption coefficient. In this way, as can be seen in Fig 5, Pb doping leads to redshift in the imaginary parts of dielectric constant, which is consistent with the electronic band structure. The single Pb doping doesn't lead to much change in the VB edges, but considerably changes the formation of the CB bottom edges. The computational results revealed that the general reason for the narrowing of the bandgap is the formation in the CB bottom edges where Pb 6s-O 2p empty energy states are dominant (see Fig. 3(b)). In pure SnO$_2$, Sn 5s-O 2p hybrid energy states are dominant in the lower parts of the CB. When Pb and Sn are replaced, Pb-O bonds are formed. In these bonds, Pb 6s energy levels replace Sn 5s and Pb 6s-O 2p hybrid energy levels form. The Pb 6s energy levels are pushing the CB bottom edges towards the band gap. Moreover, looking at the charge distributions of atoms in Table II,  the charges around the Pb atom are approximately 0.40\textit{e} more than Sn atom.  At the same time, the charge of each of the O atoms bonding with the Pb atom is about 0.06\textit{e} less than in the pure case. A possible reason is that the Pb atom having a larger ionic radius than the Sn atom, and this introduces the electron cloud to tend to overlap.

\begin{figure}[t]
    \centering
     \includegraphics[width=0.5\textwidth,trim={0 0 0 0},clip]{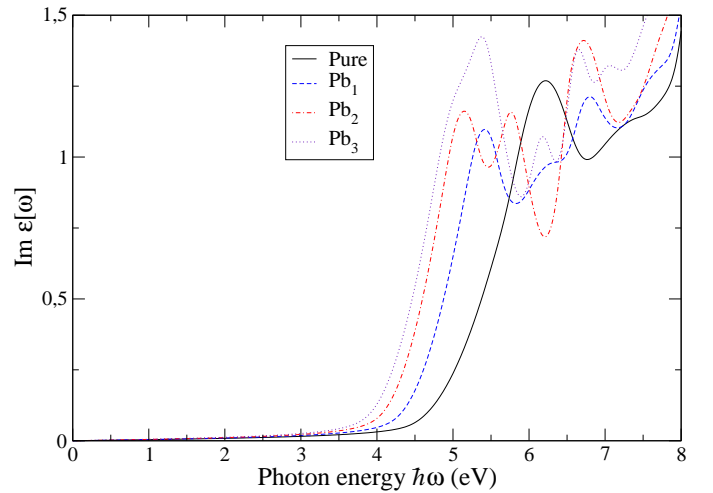}
    \caption{The imaginary part of dielectric function for pure, Pb$_1$, Pb$_2$ and Pb$_3$ structures.}
    \label{optic}
\end{figure}

\textbf{\textit{Two Pb doping(Pb$_2$)}}: The energies of various combinations were compared to find the most stable configuration of the two Pb doped SnO$_2$ (see Fig. 6). The adding of two Pb atoms into the 72-atom supercell corresponds to the impurity doping rate of  \%8.33. The measured distance between the two Pb atoms in the supercell is 3.74\AA, as will be seen in Fig. 6. Similar to that of a single Pb doping, the bond length between the Pb and the O atoms is about 2.13\AA. As can be seen in the calculated formation energies in Fig. 8, the energy of Pb$_2$ in the O-poor environment is lower than Pb$_3$ and higher than Pb$_1$, whereas in the O-rich environment the energy is lower than Pb$_1$ and higher than Pb$_3$.

\begin{figure}[h]
    \centering
    \includegraphics[width=0.3\textwidth]{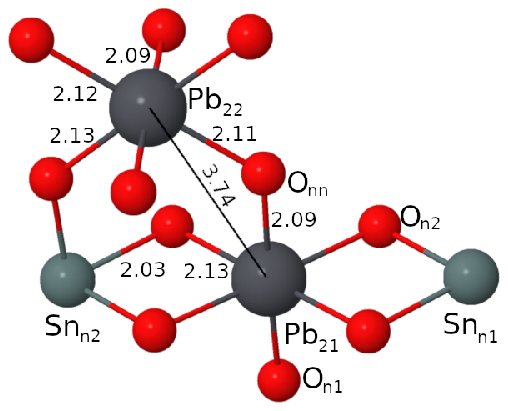}
    \caption{Relaxed structure of the two Pb dopant inside the rutile SnO$_2$. The Bond lengths are in angstroms.}
    \label{s2pb}
\end{figure}
\begin{figure}[b]
    \centering
    \includegraphics[width=0.35\textwidth]{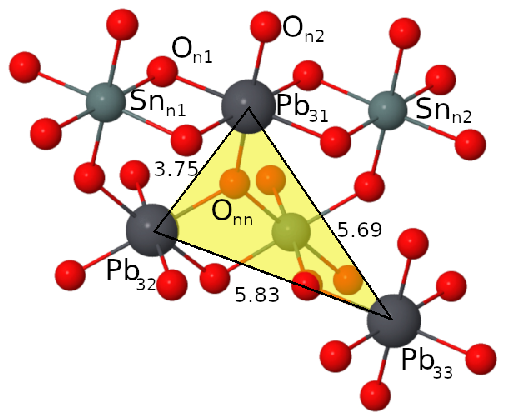}
    \caption{Relaxed structure of the three Pb dopant inside the rutile SnO$_2$. The Bond lengths are in angstroms.}
    \label{s3pb}
\end{figure}

In terms of electronic structure, as can be seen in Fig 3(c), there is no noticeable change in the shape of the VB edges. The Pb d energy states are located below the VB edge and their density increases with the addition of the second Pb compared to the Pb$_1$ case. In the VB edges O p Energy levels are dominant. On the other hand, changes occur in the CB band edge and the empty energy states start from 0.45 eV lower than the pure case. Where, unlike Pb$_1$, empty energy states occur about 0.1 eV below the CB. The O$_{nn}$ (See Fig. 6) atom which is bonded with two Pb atoms is dominant in first peak at the bottom of these separate energy states and its charge is 0.08\textit{e} less than other O atoms bonded with Pb (See Table II). The O atoms that bond with the Pb atom are dominant in the lower parts of the CB. As a result of these formations in the CB edge, the bandgap narrows to 3.19 eV, and the transitions are direct ($\Gamma \rightarrow \Gamma $). The changes in the edge of the CB are effective in the narrowing here. This narrowing is consistent with redshifting in the imaginary parts of the dielectric constant and that is more than in Pb$_1$ case (See Fig. 5). The structural and electronic properties of the double Pb doping shows similar characteristics with the one defect case, which indicates that the defect-defect interaction is weak due to the distance between the Pb atoms. 

\begin{figure}[h]
    \centering
    \includegraphics[width=0.35\textwidth]{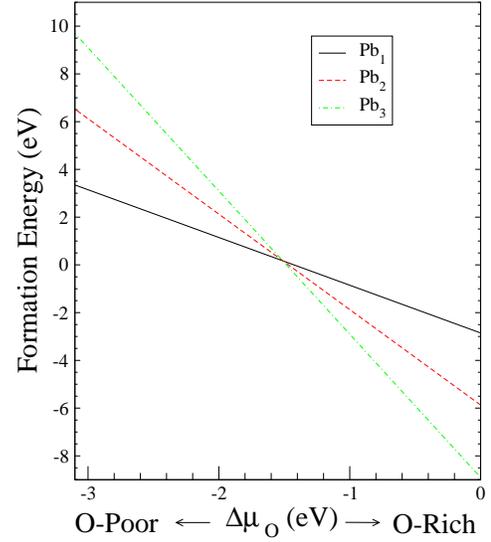}
    \caption{Calculated formation energies as a function of the oxygen chemical potential for Pb$_1$, Pb$_2$ and Pb$_3$ structures.}
    \label{f}
\end{figure}

\textbf{\textit{Three Pb doping(Pb$_3$)}}: The adding of three Pb atoms into the 72-atom supercell corresponds to the impurity doping rate of \%12.5. To find the most stable configuration of the triple Pb doping, various interatomic distances between dopants were tested by comparing the energies. Considering that the distances between Pb atoms form a triangle, as shown in Fig. 7, the lengths of Pb$_{33}$-Pb$_{31}$,  Pb$_{33}$-Pb$_{32}$ and Pb$_{31}$-Pb$_{32}$ are 5.69\AA, 5.83\AA\ and 3.75\AA, respectively. According to this structure, the third Pb atom prefers to be located farther away from the other two Pb atoms. As with the Pb$_2$ case, dopants do not tend to make a cluster. When the amount of Pb in the supercell increases, elongation occurs in the lattice parameters. In Pb$_3$ case, this elongation is 0.02\AA\ and 0.023\AA\ for \emph{a} and \emph{c}, respectively. These values are in good agreement with the experimental data.\cite{SARANGI201816} As shown in Fig. 8, the calculated formation energies of the three Pb doping under O-rich and O-poor conditions are -8.71 eV and 9.61 eV, respectively. Calculated formation energies show that under O-rich conditions, the three Pb doping are energetically more favorable than the others and vice versa under the O-poor conditions.

As can be seen in Table III, the calculation results show that with the impurity doping rate of \%12.5, the band gap of SnO$_2$ is direct ($\Gamma \rightarrow \Gamma $) and has a width of 3.02 eV. In an experimental study, \citet{SARANGI201816} show that a \%15 Pb doping narrows the band gap of SnO$_2$ to 2.98 eV. Considering that the narrowing in the band gap is proportional to the doping rate, the computational result is in good agreement with the experimental data. The band gap narrowing in Pb$_3$ case is higher than the Pb$_1$ and Pb$_2$ cases. In agreement with the experimental data, as the number of Pb increases, the band gap narrows. This feature can be used to fine-tune the width of the band gap. The obtained DOS pattern for Pb$_3$ case is shown in Fig. 3(d). As can be seen from the DOS pattern, there are no energy levels induced by dopants in the band gap. The VB band edges show similar characteristics to Pb$_1$ and Pb$_2$, and the Pb d energy levels density in VB is more than Pb$_1$ and Pb$_2$. As can be seen from the DOS pattern (see Fig. 3(d)), in the case of Pb$_3$, as in Pb$_1$ and Pb$_2$, the narrowing of the bandgap is caused by the formation of the bottom of CB, where Pb 6s-O 2p hybrid states predominate. Pb 6s-O$_{nn}$ 2p empty energy levels predominate in the first peak at the bottom of the CB. In the other two peaks, as in Pb$_2$ case, empty Pb s-O 2p hybrid energy levels induced by the other O atoms bonded with Pb predominate. As can be seen in Fig. 3(d), the addition of the third Pb atom causes the closure of the 0.1 eV wide gap between the second and third peaks in the case of Pb$_2 $. At the same time, in the case of Pb$ _3$, with the increase in the density of the peaks compared to other cases, three more intense Pb 6s-O 2p hybrid empty energy levels are formed at the bottom of the CB. In the imaginary part of the dielectric constant (see Fig. 5), there is a redshift that compatible with the narrowing of the band gap and this shifting is more than in the other cases. However, compared to other cases, the formed shoulder is higher and wider. The formation that occurs at the bottom of CB is effective in the formation of this shoulder. This result indicates that the photocatalytic activity of the Pb$_3$ case stronger in the visible region than in the other cases. 
\section{CONCLUSION}
According to the results obtained in this present work, when the LDA are used instead of PBE in HSE06 method with a mixing parameter of \%29, the electronic structure of rutile SnO$_2$ can be described as quite compatible with experimental data. On the other hand, in the Pb doped SnO$_2$ cases, using the same method, electronic structure, and especially the band gap were calculated quite compatible with the experimental data. Consistent with experimental data, the band gap narrows as the Pb doping rate increases. The calculation results obtained in this study show that the decrease in the energy level of the bottom of the CB plays an important role in the narrowing of the band gap and there is no significant change in the energy level of the top of VB. The Pb 6s energy levels are pushing the CB bottom edges towards the band gap. Pb doping doesn't induce electron-hole recombination centers in the SnO$_2$ band gap. This is important for the electron transport system. On the other hand, the narrowing in the band gap is not based on structural distortion, as Pb doping does not significantly distort the structure.  Due to this effect of the Pb atom, the energy level of the CB can be adjusted by using the doping ratio of the Pb atom and the band gap can be narrowed in a controlled manner. This is important to adjust the CB energy level of the SnO$_2$ ETL according to the type of perovskite used.
%\begin{acknowledgments}
%\end{acknowledgments}

%\section{References}

\bibliography{celik2020}

\end{document}